# A large-scale COVID-19 Twitter chatter dataset for open scientific research - an international collaboration

Juan M Banda
   Department of Computer Science, Georgia State University, Atlanta, Georgia, USA, jbanda@gsu.edu

Ramya Tekumalla
   Department of Computer Science, Georgia State University, Atlanta, Georgia, USA, rtekumalla1@gsu.edu

Guanyu Wang
   Missouri school of journalism, University of Missouri, Columbia, Missouri, gwnd6@mail.missouri.edu

Jingyuan Yu
   Department of social psychology, Universitat Autònoma de Barcelona, Barcelona, Spain, jingyuan.yu@e-campus.uab.cat

Tuo Liu
   Department of psychology, Carl von Ossietzky Universität Oldenburg, Oldenburg, Germany, tuo.liu@uol.de

Yuning Ding
   Language technology lab , Universität Duisburg-Essen, Duisburg, Germany, yuning.ding@uni-due.de

Katya Artemova
   Faculty of Computer Science, Higher School of Economics - National Research University, Moscow, Russia,, ktr.che@me.com

Elena Tutubalina
   Kazan Federal University, Kazan, Russia, eivtutubalina@kpfu.ru

Gerardo Chowell
   Department of Population Health Sciences, Georgia State University, Atlanta, Georgia, gchowell@gsu.edu

## ABSTRACT

As the COVID-19 pandemic continues its march around the world, an unprecedented amount of open data is being generated for genetics and epidemiological research. The unparalleled rate at which many research groups around the world are releasing data and publications on the ongoing pandemic is allowing other scientists to learn from local experiences and data generated in the  front lines of the COVID-19 pandemic. However, there is a need to integrate additional data sources that map and measure the role of social dynamics of such a unique world-wide event into biomedical, biological, and epidemiological analyses. For this purpose, we present a large-scale curated dataset of over 383 million tweets, growing daily, related to COVID-19 chatter generated from January 1st to June 7th at the time of writing. This open dataset will allow researchers to conduct a number of research projects relating to the emotional and mental responses to social distancing measures, the identification of sources of misinformation, and the stratified measurement of sentiment towards the pandemic in near real time.

## 1. Introduction

The ongoing COVID-19 pandemic began in the form of a cluster of viral pneumonia patients of unknown etiology in the city of Wuhan, China in December 2019. Unfortunately, the interventions to contain its spread were not implemented soon enough to limit the spread of the virus within China's borders. While transmission has been dramatically reduced in China through strict social distancing interventions, the virus was exported to multiple countries and is now generating sustained transmission in multiple areas of the world, including areas with active hotspots of the disease including the United States, Italy, Spain, and France [1] . As of November 13th, 53,305,211 cases have been recorded including 1,302,560 deaths according to the worldometer coronavirus pandemic tracker [2].

While the ongoing COVID-19 presents with unprecedented challenges to humanity, the wider scientific community can only advance science when they have access to openly available data. Social media platforms like Twitter and Facebook contain an abundance of text data that can be utilized for research purposes. Over the last decade, Twitter has proven to be a valuable resource during disasters for many-to-many crisis communication [3–5] . With Twitter data, it is possible to analyze symptom configurations, risk factors, origin, virus genetics, and spread patterns can be studied and monitored [6–9] . Recent studies [10,11] prove that data sharing improves quality and strengthens research, with collaborative efforts providing an opportunity for researchers to continually enhance research ideas and avoid redundant efforts [12,13] . We opted to release our data to the public for the greater good when the dataset accumulated 40 million tweets on March 23rd [14] . We have been providing updates every two days [15] , with a cumulative update every week, most recently on November 8th [16]. This previous update had over 800 million tweets available for researchers. The community response by word of mouth has led to over **41,592 views** and over **33,274 downloads** of the resource. Moreover, several international researchers have reached out to contribute data and provide analysis expertise. In this release we have incorporated additional data provided by our co-authors expanding the size of the dataset to over 800,064,296 tweets and added vital data on the early days of the pandemic which was unavailable during the initial data collection effort. This shows the value of this kind of data and the engagement of scientists that come together to create extensive resources for the benefit of society. Aside from providing the full dataset with retweets included, we provide a clean version with no retweets for researchers with limited resources to access a lighter version of the dataset. Furthermore, to assist researchers for NLP tasks we provide the top 1000 frequent terms, 1000 bigrams, and 1000 trigrams. The released dataset adheres with FAIR principles [17] . Due to Twitter's terms of service, tweet text cannot be shared. Therefore, tweet ids are publicly made available using Zenodo [18]. The tweet ids can be hydrated using tools like Social Media Mining Toolkit or twarc [19,20] . The deliverables [15,18] include tweet ids and code to process the tweets. Please note that the code to process the tweets would work only after the tweets are hydrated. We provided the date and time meta-data elements on our dataset for groups wanting to target their research questions to certain days to avoid having to hydrate the whole resource at once. We have automated pipelines to continue collecting tweets as the pandemic runs its course and to provide updates every two days on our Github repository. We are also welcoming any additional data that provides new tweets to our resource.

## 2. Methods

The initial versions of this dataset [14,21] only included data collected from the publicly available Twitter Stream API with a collection process that gathered any available tweets within the daily restrictions from Twitter from January to March 11th, filtering them on the following 3 keywords: "coronavirus", "2019ncov", "corona virus". We shifted our focus to collect exclusively COVID-19 tweets on March 12th, 2020 with the following keywords: "COVD19", "CoronavirusPandemic", "COVID-19", "2019nCoV", "CoronaOutbreak", "coronavirus", "WuhanVirus", thus the number of tweets gathered dramatically expanded the dataset. Please note that the Stream API only allows free access to a one percent sample of the daily stream of Twitter. Our methodology relies on Python and the Tweepy package [22] , as in our previous work [23]. We recently received another set of 30+ Million tweets collected from January 27th, 2020 to March 27th, 2020 from our co-author, Jingyuan Yu, and his collaborators with the following keywords: "coronavirus", "wuhan", "pneumonia", "pneumonie", "neumonia", "lungenentzündung", "covid19". These tweets were collected in the following languages: English, French, Spanish, and German, while our original collection is done for any language available. We have fully integrated and deduplicated our collaborators' tweet collection with ours, thus the numbers and tweets presented in this dataset are of unique tweet identifiers from January 1st to November 8th (at the time of writing). In version 10 we added ~1.5 million tweets in the Russian language collected between January 1st and May 8th, gracefully provided to us by our co-authors Katya Artemova and Elena Tutubalina. Table 1 represents the monthly number of tweets included in this dataset.

**Table 1: Number of Covid-19 chatter tweets in this dataset**

| Month | Full | Clean |
| --- | --- | --- |
| January | 6,737,966 | 1,329,483 |
| February | 27,666,656 | 5,886,751 |
| March | 111,006,589 | 21,612,183 |
| April | 128,048,263 | 31,661,550 |
| May | 120,186,704 | 31,361,965 |
| June | 92,566,134 | 23,410,940 |
| July | 97,185,376 | 23,595,378 |
| August | 73,931,454 | 18,614,572 |
| September | 60,905,104 | 15,297,347 |
| October | 71,329,206 | 18,009,735 |
| November (as Nov 8th) | 10,500,844 | 3,492,272 |
| **Total** | **800,064,296** | **194,272,176** |

As previously mentioned, the number of collected tweets increased tremendously since starting dedicated collection. All our preprocessing scripts utilize components of the Social Media Mining Toolkit (SMMT) [24] . We make a distinction between our full and clean versions of the dataset. The full dataset consists of both tweets and retweets. There are several practical reasons to leave the retweets; tracing important tweets and their dissemination is one of them. A clean version with no retweets is also released, intended for NLP researchers. We also release extracted frequent terms, bigrams, and trigrams for this community. Figure 1 outlines the steps taken to build our dataset.

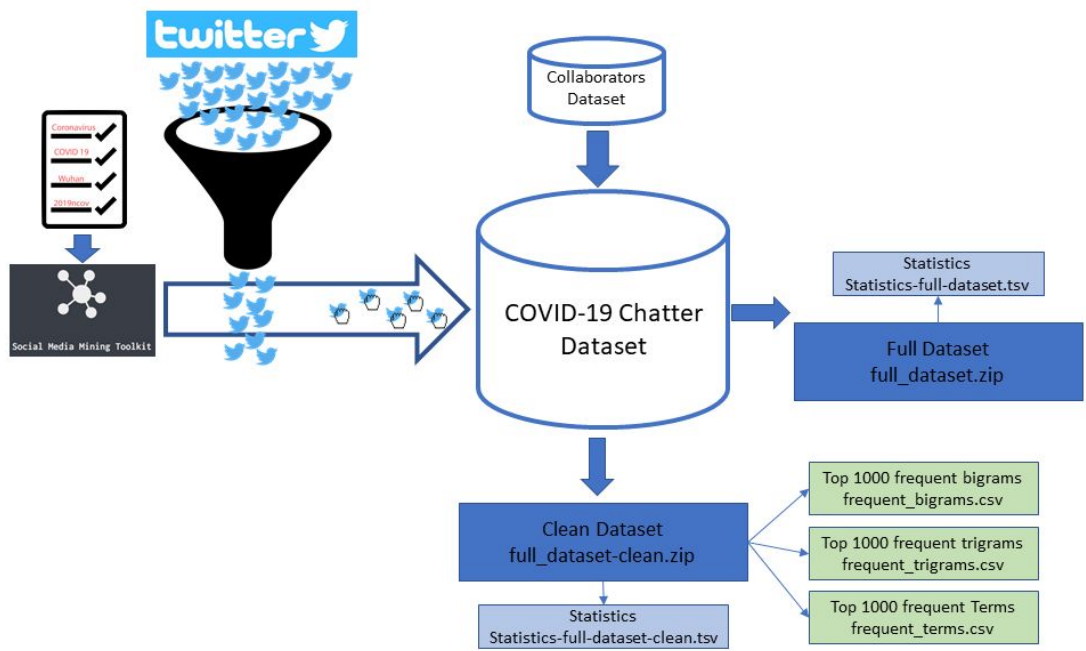

**Figure 1: Dataset gathering and construction steps.**

As shown in Figure 1, we used SMMT to listen to the Twitter Stream API for tweets with the described keywords. We then gather all the tweets that have the desired keywords before aggregating them locally. Our contributors used a similar procedure to gather their tweets and provided us with tab delimited files with their data. We processed them to fit our own local format to be able to include them in our dataset after deduplication (removal of tweets we have in common) and only keep unique tweet identifiers between the datasets. We then preprocess the large set of tweets to extract the shareable meta-data of the full dataset (tweet_id, collected date, collected time), preparing the full_dataset.tsv.gz file. At the same time we also remove tweets that are re-tweeted (this is, existing tweets that are re-shared by others) to create the full_dataset-clean.tsv.gz file. Our preprocessing involves cleaning up special characters, such as carriage returns, removing urls and large blank spaces. Our preprocessing is rather relaxed as we are leaving all available languages intact. To generate the frequent terms and ngrams (sets of n-terms that appear constantly together), we remove all stop words in English and Spanish, using the Spacy [25] . These lists are originally quite large, so we only share the top 1000 terms, bigrams, and trigrams. We continue to update our original dataset every two days [14] with major releases every week [18,21] and plan to continue doing this for at least the next 12 months, a period that will likely cover the main pandemic period.

## 3. Data Validation and Control

The dataset is made available through Zenodo [18]. There are 7 files in this repository. Table 2 details the files, formats, and their utility. The example column consists of a sample line from the files. The tweet ids in the dataset can be hydrated using SMMT. The hydrated tweets would produce a JSON object for each tweet id. It is important to note that when users remove their accounts or individual tweets, these get removed and are no longer available for download. In such cases, we can share the data on request while adhering to the Twitter data sharing policy. The frequent terms, bigrams, and trigrams are retrieved from the cleaned version of the dataset. The full_dataset.tsv consists of all the procured tweet ids. The full_dataset-clean.tsv contains only original tweets with no retweets. While some applications and questions are better served with the full dataset, NLP researchers might prefer a clean dataset to have less inflated counts of the n-grams identified.

**Table 2: Details of released COVID-19 dataset. Note that the word TAB is not found, but instead use the special '\t' character for this. We are showing it on the descriptions for illustrative purposes.**

| File Name | Description | Example |
|---|---|---|
| full_dataset.tsv.gz | A zipped, tab separated file which contains all the tweet ids in the format - Tweet ID TAB Date TAB Time TAB language TAB country_code | 1238315928095297538 TAB 2020-3-13 TAB 4:8:18 TAB en TAB US |
| full_dataset-clean.tsv.gz | A zipped, tab separated file which does not contain any retweet ids in the format - Tweet ID TAB Date Tab Time | 1238315936379293696 TAB 2020-3-13 TAB 4:8:20 |
| statistics-full_dataset-clean.tsv | A tab separated file which contains counts of total tweets each day for the clean dataset in the format - Date TAB Total No of Tweet Ids | 2020-3-13 TAB 751804 On March 13,2020 a total of 751,804 clean tweets were collected |
| statistics-full_dataset.tsv | A tab separated file which contains counts of total tweets each day for full dataset in the format - Date TAB Total No of Tweet Ids | 3/13/2020 TAB 4160194 On March 13,2020 a total of 4,160,194 tweets were collected |
| frequent_terms.csv | A comma separated file which contains the counts of top 1000 frequent terms in the following format - term, Total count | covid19, 1767060 covid 19 term appeared in 1,767,060 tweets |
| frequent_bigrams.csv | A comma separated file which contains counts of top 1000 bigrams in the format - gram, Total count | covid 19, 1467434 covid 19 bigram appeared in 1,467,434 tweets. |
| frequent_trigrams.csv | A comma separated file which contains counts of top 1000 trigrams in the format - gram, Total count | coronavirus covid 19, 52143 coronavirus covid 19 appeared in 52,143 tweets |
| emoji.zip | A zipped collection of dated files which contain the top emojis, both in text and unicode character versions, and their frequencies per day for all clean tweets. | face with tears of joy TAB 3651 loudly crying face TAB1591 |
| hashtag.zip | A zipped collection of dated files which contain the top hashtags and their frequencies per day for all clean tweets. | #socialdistancing TAB 2327 #staysafe TAB 2056 |
| mentions.zip | A zipped collection of dated files which contain the top mentions (@) and their frequencies per day for all clean tweets. | @whitehouse TAB 1415 @borisjohnson TAB 1325 |

## 4. Re-use Potential

In order to use our resource, we have provided all the software tools we utilized to preprocess, clean and parse the Twitter data on our Github repository [15] , under the processing code directory. Note that the tweets need to be hydrated first using tools like Social Media Mining Toolkit or twarc [19,20] . Once the tweets are hydrated and a JSON object has been returned, we use the files **parse_json_extreme.py** and **parse_json_extreme_clean.py** to extract the tweet identifier, date of creation, text, language and a few other extra fields. This process can be configured by adding which fields from the tweet json object the user wants to extract in the fields.py file. These utilities produce a full and a clean version of the dataset respectively, on a tab delimited file. This process is optimized to read large files without loading them fully in memory. If the user has a system with very large amounts of RAM memory, we also provide **parse_json_lite.py** to perform the same task. Once the JSON object has been parsed, most users will be able to operate on the tweets directly this way. We additionally provide the **get_1grams.py** and **get_ngrams.py** utilities to generate the most frequent terms and bigrams and trigrams, respectively. As the hydrated tweet JSON objects are typically quite large, we recommend separating them in daily batches to be able to more efficiently process them. All our previously mentioned tools take a single file as an input parameter for processing and output a new file. In order to combine the results of the ngram generation from multiple files, we prove the following tools that take a folder path as input and iterate through all files present: **combine1grams.py, combineNgrams.py**. In order to share the tweet identifiers with other groups, we provide the getDataset.py, getDataset_clean.py files which generate the equivalent files of **full_dataset.tsv** and **full_dataset-clean.tsv** that are presented in this resource in a compressed (zip) maner. Lastly, dataset statistics can be calculated with **getStats.py**, by passing the full or clean dataset filename to them.

### ACKNOWLEDGEMENT

This work was partially supported by the National Institute of Aging through Stanford University's Stanford Aging & Ethnogeriatrics Transdisciplinary Collaborative Center (SAGE) center (award 3P30AG059307-02S1).

### REFERENCES

1.  World Health Organization. WHO characterizes COVID-19 as a pandemic. [cited 27 Mar 2020]. Available: https://www.who.int/emergencies/diseases/novel-coronavirus-2019/events-as-they-happen

2. Coronavirus Update (Live): 737,575 Cases and 34,998 Deaths from COVID-19 Virus Outbreak - Worldometer. [cited 30 Mar 2020]. Available: https://www.worldometers.info/coronavirus/

3. Bruns A, Liang YE. Tools and methods for capturing Twitter data during natural disasters. First Monday. 2012;17: 1–8. Available: http://eprints.qut.edu.au/49716

4. Zou L, Lam NSN, Cai H, Qiang Y. Mining Twitter Data for Improved Understanding of Disaster Resilience. Ann Assoc Am Geogr. 2018;108: 1422–1441. doi:10.1080/24694452.2017.1421897

5. Earle P. Earthquake Twitter. Nat Geosci. 2010;3: 221–222. doi:10.1038/ngeo832

6. Gao J, Tian Z, Yang X. Breakthrough: Chloroquine phosphate has shown apparent efficacy in treatment of COVID-19 associated pneumonia in clinical studies. Biosci Trends. 2020;14: 72–73. doi:10.5582/bst.2020.01047

7. Xu Z, Shi L, Wang Y, Zhang J, Huang L, Zhang C, et al. Pathological findings of COVID-19 associated with acute respiratory distress syndrome. Lancet Respir Med. 2020. doi:10.1016/S2213-2600(20)30076-X

8. Zhou F, Yu T, Du R, Fan G, Liu Y, Liu Z, et al. Clinical course and risk factors for mortality of adult inpatients with COVID-19 in Wuhan, China: a retrospective cohort study. Lancet. 2020;395: 1054–1062. doi:10.1016/S0140-6736(20)30566-3

9. Tekumalla R, Banda JM. Characterization of Potential Drug Treatments for COVID-19 using Social Media Data and Machine Learning. 2020. Available: https://arxiv.org/abs/2007.10276

10. Warren E. Strengthening Research through Data Sharing. N Engl J Med. 2016;375: 401–403. doi:10.1056/NEJMp1607282

11. Saez-Rodriguez J, Costello JC, Friend SH, Kellen MR, Mangravite L, Meyer P, et al. Crowdsourcing biomedical research: leveraging communities as innovation engines. Nat Rev Genet. 2016;17: 470–486. doi:10.1038/nrg.2016.69

12. Emmert-Streib F, Dehmer M, Yli-Harja O. Against Dataism and for Data Sharing of Big Biomedical and Clinical Data with Research Parasites. Front Genet. 2016;7: 154. doi:10.3389/fgene.2016.00154

13. Greene CS, Garmire LX, Gilbert JA, Ritchie MD, Hunter LE. Celebrating parasites. Nature genetics. 2017. pp. 483–484. doi:10.1038/ng.3830

14. Banda JM, Tekumalla R. A Twitter Dataset of 40+ million tweets related to COVID-19. 2020. doi:10.5281/zenodo.3723940

15. Banda Juan M, Tekumalla Ramya. Covid-19 Twitter dataset and pre-processing scripts. [cited 27 Mar 2020]. Available: https://github.com/thepanacealab/covid19_twitter

16. Banda JM, Tekumalla R, Wang G, Yu J, Liu T, Ding Y, et al. A large-scale COVID-19 Twitter chatter dataset for open scientific research - an international collaboration. 2020. doi:10.5281/zenodo.3723939.

17. Wilkinson MD, Dumontier M, Aalbersberg IJJ, Appleton G, Axton M, Baak A, et al. The FAIR Guiding Principles for scientific data management and stewardship. Sci Data. 2016;3: 160018. doi:10.1038/sdata.2016.18

18. Banda JM, Tekumalla R, Wang G, Yu J, Liu T, Ding Y, et al. A Twitter Dataset of 383+ million tweets related to COVID-19. 2020. doi:10.5281/zenodo.3884334

19. Tekumalla R, Banda JM. Social Media Mining Toolkit (SMMT). Genomics Inform. 2020;18: e16. doi:10.5808/GI.2020.18.2.e16

20. twarc. Github; Available: https://github.com/DocNow/twarc

21. Banda JM, Tekumalla R, Chowell G. A Twitter Dataset of 70+ million tweets related to COVID-19. 2020. doi:10.5281/zenodo.3732460

22. tweepy. Github; Available: https://github.com/tweepy/tweepy

23. Tekumalla R, Asl JR, Banda JM. Mining Archive. org's Twitter Stream Grab for Pharmacovigilance Research Gold. Proceedings of the International AAAI Conference on Web and Social Media. 2020. pp. 909–917. Available: https://www.aaai.org/ojs/index.php/ICWSM/article/view/7357

24. Tekumalla R, Banda JM. Social Media Mining Toolkit (SMMT). arXiv [cs.IR]. 2020. Available: http://arxiv.org/abs/2003.13894

25. Explosion AI. spaCy-Industrial-strength Natural Language Processing in Python. 2017. Available: https://spacy. io